\documentclass[aps,twocolumn,groupedaddress,amsmath,amssymb]{revtex4-2}
\usepackage{graphicx}
\usepackage{color}

\begin{document}

\title{Towards the emergence of non-zero thermodynamical quantities for Lanczos-Lovelock black holes dressed with a scalar field
}

\author{Mois\'es Bravo-Gaete}
\email{mbravo-at-ucm.cl} \affiliation{Facultad de Ciencias
B\'asicas, Universidad Cat\'olica del Maule, Casilla 617, Talca,
Chile.}

\author{Carlos G\'omez Gaete}
\email{cgomez-at-ucm.cl} \affiliation{Facultad de Ciencias
B\'asicas, Universidad Cat\'olica del Maule, Casilla 617, Talca,
Chile.}

\author{Sebasti\'an G\'omez Rodr\'iguez}
\email{sebastian.gomez-at-uautonoma.cl} \affiliation{Facultad de Ingenier\'ia, Universidad Aut\'onoma de Chile, 5 poniente 1670, Talca, Chile.}

\author{Luis Guajardo}
\email{lguajardo-at-ucm.cl} \affiliation{Facultad de Ciencias
B\'asicas, Universidad Cat\'olica del Maule, Casilla 617, Talca,
Chile.}

\begin{abstract}
The present work aims to explore the model given by Lanczos-Lovelock gravity theories indexed by a fixed integer to require a unique anti-de Sitter vacuum, dressed by a scalar field non-minimal coupling. For this model, we add a special matter source characterized by a non-linear Maxwell field coupling with a function depending on the scalar field. Computing its thermodynamics parameters by using the Euclidean action, we obtain interesting and non zero thermodynamical quantities, unlike its original version, allowing analyzing thermodynamical stability. Together with the above, we found that these solutions satisfy the First Law of Thermodynamics as well as a Smarr relation.
\end{abstract}

\maketitle
\newpage

\section{Introduction}

It is well known that the most general gravity theory in four dimensions is given by the Einstein-Hilbert action together with a cosmological constant, where its dynamic is codified in non-linear second order differential equations. Nevertheless, motivations such as the accelerated expansion of the Universe \cite{Riess:1998cb}, as well as the first detection of gravitational waves \cite{TheLIGOScientific:2017qsa,Monitor:2017mdv} have motivated the active exploration of modified theories of gravity, being higher dimensional gravity theories an interesting object of study, in particular in the context of the Anti-de Sitter/Conformal Field Theory  duality \cite{Maldacena:1997re}. A via of exploration considering this approach is given  by the introduction of higher powers of the curvature terms. Under this scenario, we have the Lanczos-Lovelock action \cite{Lanczos:1938sf,Lovelock:1971yv}, corresponding to the most general action in higher dimensions such that the equations of motions with respect to the metric are at most of second-order, given by
\begin{eqnarray}\label{lovelock}
&&\int \sum_{p=0}^{[D/2]}\alpha_p~ L^{(p)},\\
&& L^{(p)}=\epsilon_{a_1\cdots a_d} R^{a_1a_2} \wedge \cdots \wedge
R^{a_{2p-1}a_{2p}}\wedge e^{a_{2p+1}}\wedge \cdots  \wedge e^{a_d}.\nonumber
\end{eqnarray}
Here (\ref{lovelock}) is represented as a polynomial with a degree
$[ D/2 ]$,  wherein our notations $[ \,]$ is the integer part, and is written in term of the Riemann curvature $R^{ab} = d\,\Sigma^{ab} + \Sigma^{a}_{\;c} \wedge
\Sigma^{cb}$ and the vielbein $e^{a}$, together with the coupling constant  $\alpha_{p}$. About their general properties and solutions see for example the references \cite{Charmousis:2008kc,Garraffo:2008hu,Camanho:2011rj}. For topological AdS black holes and their thermodynamical atributes see the references \cite{Banados:1993ur,Cai:1998vy,Crisostomo:2000bb,Aros:2000ij,Arenas-Henriquez:2019rph}.

Together with the above, as was shown in \cite{Crisostomo:2000bb}, to construct a theory with a unique AdS vacuum, fixing the cosmological constant, the coupling constants $\alpha_{p}$'s must be tied and yielding a set of theories indexed through to an integer $n$ which reads
\begin{eqnarray}
&&S_{(n)}=-\frac{1}{2n(D-3)!  }\int \sum_{p=0}^{n}
{n \choose p }\frac{L^{(p)}}{(D-2p)},\\ \label{Ik}
&&1\leq n\leq \Big[\frac{D-1}{2}\Big], \nonumber
\end{eqnarray}
where
$$
{n \choose p }=
\frac{n!}{p!\, (n-p)!},$$
and the action (\ref{Ik}) can be recast as
\begin{eqnarray}
I_{(n)}&=&\frac{1}{2}\int d^{D}x \,\sqrt{-g} \mathcal{L}_{(n)}(g_{\mu \nu}, R_{\mu \nu \sigma \rho}),\label{Ik2} \\
&=&\frac{1}{2}\int d^{D}x\,\sqrt{-g}
\Big[R+\frac{(D-1)(D-2)}{n}
\nonumber\\
&+&\frac{(n-1)}{2!(D-3)(D-4)}{L}^{(2)}\nonumber\\
&+& \frac{(n-1)(n-2)}{3!(D-3)(D-4)(D-5)(D-6)}{L}^{(3)}+\cdots \Big],\nonumber
\end{eqnarray}
where $R$ is the scalar curvature and
\begin{eqnarray*}
L^{(2)}&=&R^{2}-4\,R_{\mu \nu}R^{\mu
\nu}+R_{\alpha\beta\mu\nu}R^{\alpha\beta\mu\nu},\label{gaussbonnet}\\
L^{(3)}&=&R^3   -12RR_{\mu \nu } R^{\mu \nu } + 16\,R_{\mu \nu
}R^{\mu }_{\phantom{\mu} \rho }R^{\nu \rho }\nonumber\\
&+& 24 R_{\mu \nu
}R_{\rho \sigma }R^{\mu \rho \nu \sigma }
+ 3RR_{\mu \nu \rho \sigma
} R^{\mu \nu \rho \sigma }\nonumber\\
&-&24R_{\mu \nu }R^{\mu} _{\phantom{\mu} \rho \sigma \kappa } R^{\nu
\rho \sigma \kappa  }+ 4 R_{\mu \nu \rho \sigma }R^{\mu \nu \eta
\zeta } R^{\rho \sigma }_{\phantom{\rho \sigma} \eta \zeta }\nonumber\\
&-&8R_{\mu
\rho \nu \sigma } R^{\mu \phantom{\eta} \nu \phantom{\zeta}
}_{\phantom{\mu} \eta \phantom{\nu} \zeta } R^{\rho \eta \sigma
\zeta }.\label{cubiclagrangian}
\end{eqnarray*}

On the other hand, the addition of scalar field $\Phi$ non-minimally coupled to the scalar curvature $R$ and a potential $U(\Phi)$ as a matter source
\begin{eqnarray}
S_{\Phi}&=&\int {d}^D x\sqrt{-g} \Biggl[
-\frac{1}{2}\nabla_{\mu}\Phi\nabla^{\mu}\Phi-\frac{\xi}{2}R\Phi^2-U(\Phi)\Biggr]\nonumber\\
&=&\int {d}^D x\sqrt{-g}  \mathcal{L}_{\Phi},\label{eq:LPhi}
\end{eqnarray}
allow to obtain a variety of black holes solutions with a planar base manifold for the event horizon with a wide range of values for the non-minimally coupled parameter $\xi$ \cite{Correa:2013bza}. Just for completeness, this kind of matter source has been a good toy model to find non-minimally dressed black holes solutions in three
\cite{Martinez:1996gn,Henneaux:2002wm,Bravo-Gaete:2020ftn}, four \cite{Bocharova:1970skc,Bekenstein:1974sf,Martinez:2005di,Cisterna:2019uek,Anabalon:2012tu,Cisterna:2021xxq} and higher dimensions \cite{Correa:2013bza,Gaete:2013ixa,Gaete:2013oda,Erices:2017izj,Ayon-Beato:2019kmz}, with different gravity theories and matter sources, even black holes solutions with non standard asymptotically behaviour.

As it was argued in \cite{Correa:2013bza} the thermodynamics analysis of these configurations shows that the mass as well as the entropy vanish, which makes the unique integration constant arose from the configuration to be treated as a sort of hair. The present work aims to explore this issue by considering a more general model that includes a non-linear Maxwell source coupled to the scalar field $\Phi$. In doing so, and as we will show below, the indicated integration constant is no longer a hair, making possible a nontrivial thermodynamics analysis for the black hole solutions in a certain range for the non-minimal coupling $\xi$. Moreover, we exhibit a certain limit in the proposed theory that correctly reproduces the results from \cite{Correa:2013bza}.

The plan of the paper is organized as follows: In Section \ref{Section-sol}, we will present the procedure elaborated in \cite{Correa:2013bza} and we will propose that with a suitable structure for the metric function as well as an addition of a matter source, we can construct a non-minimally dressed charged black hole with non-null thermodynamical quantities. In Section \ref{Section-soln} we will present an electrically charged configuration while in Section \ref{Section-term} we explore its non-null thermodynamic properties. Finally, the Section \ref{Section-conclusions}  is devoted to our conclusions and discussions.

\section{Action, field equations and derivation of the hairy solution}\label{Section-sol}

In order to be as clear and self-contained as possible, we will start deriving the solution reported in \cite{Correa:2013bza}, considering the action
\begin{eqnarray}\label{eq:actioncorrea}
S=I_{(n)}+S_{\Phi},
\end{eqnarray}
where $I_{(n)}$ and $S_{\Phi}$ are given by the gravity (\ref{Ik2}) and matter source (\ref{eq:LPhi}) contribution respectively. The field equations with respect to the metric $g_{\mu \nu}$ and the scalar field $\Phi$ read
\begin{eqnarray}
{{\cal E}}^{(n)}_{\mu\nu}:={{\cal G}}^{(n)}_{\mu\nu}- T_{\mu \nu}^{\Phi}=0,\label{eq:gmunu}\\
{{\cal E}}_{\Phi}:=\Box \Phi-\xi R\Phi-\frac{dU(\Phi)}{d\Phi}=0,\label{eq:phi}
\end{eqnarray}
with
\begin{eqnarray*}
{\cal G}^{(n)}_{\mu\nu}&=& P_{( \mu}^{\alpha \beta \gamma} R_{\nu) \alpha \beta \gamma }- 2 \nabla^{\rho} \nabla^{\sigma} P_{\mu \nu \sigma \rho}-\frac{1}{2} g_{\mu \nu} \mathcal{L}_{(n)},\\
T_{\mu \nu}^{\Phi}&=&\nabla_{\mu}\Phi\nabla_{\nu}\Phi - g_{\mu\nu}\Bigl[\frac{1}{2}\nabla_{\sigma}\Phi\nabla^{\sigma}\Phi +U(\Phi)\Bigr]\nonumber \\
&+& { \xi(g_{\mu\nu}\Box -\nabla_{\mu} \nabla_{\nu}+G_{\mu\nu} )\Phi^2},\\
\end{eqnarray*}
where $P^{\mu \nu \sigma \rho}=\delta \mathcal{L}_{(n)} / \delta R_{\mu \nu \sigma \rho}$ with $\mathcal{L}_{(n)}$ the lagrangian given previously in (\ref{Ik2}). The metric Ansatz of our study takes the form
\begin{eqnarray}\label{metric}
ds^2=-r^2\big(1-f(r)\big) dt^2+\frac{dr^2}{r^2\big(1-f(r)\big)}+r^2 \sum_{i=1}^{D-2} dx_{i}^{2},\nonumber\\
\end{eqnarray}
while that $\Phi=\Phi(r)$. For the sake of completeness, the equations of motion with respect to the metric and the potential $U(\Phi)$ are reported in the appendix.

With all these components, we start the analysis by using the combination $\mathcal{E}_{t}^{t}-\mathcal{E}_{r}^{r}=0$, yielding a second-order differential equation for the scalar field given by
$$2r^2 \left[\xi \Phi \Phi''+\left(\xi-\frac{1}{2}\right)
(\Phi')^2\right](1-f)=0,$$
where $(')$ denotes the derivative with respect the radial coordinate $r$. For $f \neq 1$, the scalar field takes the form
\begin{equation}\label{eq:phixi}
\Phi(r)=(ar-b)^{\frac{2\xi}{4 \xi-1}},
\end{equation}
and to satisfy $\displaystyle {\lim_{r \rightarrow +\infty} \Phi = 0}$ we need \begin{equation}\label{eq:rangexi}
0<\xi<\frac{1}{4},
\end{equation}
while that for the particular case $\xi=1/4$ the scalar field is given as follows
\begin{equation}\label{eq:phixi14}
\Phi(r)=b e^{ar},
\end{equation}
where for both cases $a$ is a possitive integration constant and $b$ is a positive parameter presents in the potential $U(\Phi)$. Then, under the substitution
\begin{equation}
\Sigma=f^{n-1}-  \xi \Phi^{2},
\end{equation}
the combination
\begin{equation}\label{eq:Eii}
\mathcal{E}_{t}^{t}-\mathcal{E}_{x_i}^{x_i}
=(\Sigma f' r^{D})'=0
\end{equation}
is trivially satisfied if $\Sigma=0$ and $n>1$, implying that
\begin{equation}
f=( \xi \Phi^{2})^{\frac{1}{n-1}}, \label{eq:finitial}
\end{equation}
keeping the same range for $\xi$ given in  (\ref{eq:rangexi}), where $\displaystyle {\lim_{r \rightarrow +\infty} f =0}$ is satisfied. Finally, one checks that the potential $U(\Phi)$ is precisely the one who supports the remaining Einstein equation, $\mathcal{E}_{t}^{t}=0$, as well as the scalar field equation (\ref{eq:phi}).

Although this model allows a family of black hole configurations for a wide range of values for the non-minimally coupled parameter $\xi$, at the time to study their thermodynamic properties, the mass as well as the entropy of the solution vanish trivially, and the integration constant $a$ from the scalar field can be viewed as a sort of hair \cite{Correa:2013bza}. The clue of this conclusion is in the fact that the entropy of the solution is proportional to
\begin{eqnarray} \label{Sprop}
\left(1-\xi \Phi^{2}\right)\big{|}_{r_h},
\end{eqnarray}
which is zero when the metric function takes the form (\ref{eq:finitial}), where $r_h$ is the location of the event horizon for $f$.

A way to improve this solution is by constructing a more general model, looking for non-vanishing thermodynamic parameters. This, in turn, is devoted to the following section, when we will analyze an electrical configuration.

\section{Electrically charged hairy black hole}\label{Section-soln}

As we saw above, one of the main issues is that the entropy is proportional to a quantity that vanishes at the horizon. In order to avoid this, a possible idea is to add new sources that perturb the geometry of the solution. To that end, for reasons that will be clear below, let us consider the addition of two parameters $\eta$ and $q$ into the self-interacting potential $U(\Phi)$, now reading
\begin{eqnarray}
U_{\eta,q}(\Phi)&=& \frac{1}{(1-4\xi)^2}\sum_{i=1}^{6} \beta_{i} \Phi^{\gamma_{i}},\label{eq:potxi}
\end{eqnarray}
where
\begin{eqnarray*}
\beta_{1}&=&\frac{[4(D-1)\xi-D+2]
[4\xi D-D+1]\xi}{2},\\
\gamma_1&=&2,\\ \\
\beta_{2}&=&{4[4(D-1)\xi-D+2]b\xi^2},\qquad \gamma_2=\frac{1}{2 \xi},\\ \\
\beta_3&=&{2\xi^2 b^2},\qquad \gamma_3=\frac{1-2\xi}{\xi},\\ \\
\beta_4&=& \Big\{\eta ^{\frac{1}{n-1}}\,
[4 n D \xi-(n+4\xi-1)(D-1)]\\
&\times&
\big[\big(-4nq(-n+Dn-D+2)+2n\big)\xi^2\\
&+&\big(q (n^2D-4n \eta+2 n-D n+8 \eta+4Dn\eta-2n^2\\
&-&4\eta D)
-2n \eta\big)\xi-\eta q (n-1) (D-2)\big]\Big\}\\
&\Big{/}&\big\{2 q n (n-1)^2\big\},\\
\gamma_4&=&\frac{2n}{n-1},\\ \\
\beta_5&=& \big\{\eta^{\frac{1}{n-1}}\xi b \big[\big(-16nq(-n+Dn-D+2)+4D n\\
&+&8-4D\big)\xi^2+\big((4n(Dn-2n-D)
+8n\\
&+&16\eta(-n+Dn-D+2))q
-4\eta (Dn+2-D)\\
&-&(n-1)(D-2)\big)\xi-4\eta q (n-1)(D-2)
\\
&+&\eta (n-1)(D-2)\big]\big\}\Big{/}\big\{q(n-1)^2\big\},\\
\gamma_5&=&\frac{4 \xi+n-1}{2 \xi (n-1)},
\end{eqnarray*}
\begin{eqnarray*}
\beta_6&=& \frac{\eta^{\frac{1}{n-1}}b^2(n+4\xi-1)\xi [\eta(2q-1)-\xi(2nq-1)]}
{q (n-1)^2},\\ 
\gamma_6&=&\frac{(4-2n)\xi+n-1}{\xi(n-1)}.
\end{eqnarray*}
It can be observed that the parameter $\eta$ reproduces the potential $U(\Phi)$ from \cite{Correa:2013bza} in the limit $\eta \to \xi$. On the other hand, the parameter $q$ corresponds to the power of a non-linear Maxwell source, coupled to the scalar field $\Phi$ and represented by the action
\begin{eqnarray}\label{eq:maxwell}
S_{M}&=&-\frac{1}{4}\int{d}^Dx\sqrt{-g}
\epsilon (\Phi) \left(F_{\mu \nu} F^{\mu \nu}\right)^{q}
\end{eqnarray}
with $F_{\mu \nu}:=\partial_{\mu} A_{\nu}-\partial_{\nu} A_{\mu}$, $A_{\mu} dx^{\mu}=A_{t}(r)dt$. Seminal works of a nonlinear electrodynamics source can be found in  \cite{Gonzalez:2009nn,Maeda:2008ha,Hassaine:2008pw}.  Since we are interested in real solutions, we will restrict $q$ to be a nonzero rational number with odd denominator. The coupling function $\epsilon (\Phi)$ is given by
\begin{eqnarray}
\epsilon(\Phi)^{\frac{1}{2q-1}}&=&
\left\{\Phi^{\frac{1-n+2\xi(n-2)}{\xi(n-1)}} \left(\Phi^{\frac{4\xi-1}{2\xi}}+b\right)^{\frac{1+2(1-D)q}{2q-1}}\right\}\nonumber\\
&\Big{/}&\Big\{\left[4(1+(n-1)D)\xi-(n-1)(D-1)\right] \Phi^{\frac{4\xi-1}{2\xi}}\nonumber\\
&+&b(n+4\xi-1)\Big\}.\label{eq:epsilon}
\end{eqnarray}
In resume, we are interested in the following action:
\begin{eqnarray}
S &=& \int d^{D}x\sqrt{-g} \left[ \dfrac{\mathcal{L}_{(n)}}{2} -\frac{1}{2}\nabla_{\mu}\Phi\nabla^{\mu}\Phi \right.\nonumber \\
&-&\left.\frac{\xi}{2}R\Phi^2 - U_{\eta,q}(\Phi)-\dfrac{1}{4}\epsilon (\Phi) (F_{\mu \nu} F^{\mu \nu})^{q} \right].
\end{eqnarray}
Using (\ref{eq:gmunu})-(\ref{eq:phi}), and  replacing $U(\Phi) \mapsto U_{\eta,q}(\Phi)$ {accordingly, the} equations of motions for the complete model now read
\begin{subequations}\label{eq:EOM}
\begin{eqnarray}
&&E_{\mu \nu}:={{\cal E}}^{(n)}_{\mu\nu}- T_{\mu\nu}=0, \label{eq:Einstein_phi_M}\\
&& {{\cal E}}_{\Phi}-\frac{1}{4} \frac{d \epsilon}{d \Phi} \left( F_{\alpha \beta} F^{\alpha \beta}\right)^{q}=0\label{eq:phi_M},\\
&& \nabla_{\mu} \Big(\epsilon(\Phi) \left( F_{\alpha \beta} F^{\alpha \beta}\right)^{q-1} F^{\mu \nu}\Big)=0,\label{eq:M}
\end{eqnarray}
where
\begin{equation*}
T_{\mu \nu}=\epsilon(\Phi)\Big[q \left(F_{\sigma \rho} F^{\sigma \rho}\right)^{q-1} F_{\mu \sigma} F_{\nu}^{\phantom{\nu}\sigma}-\frac{1}{4} g_{\mu \nu}\left(F_{\sigma \rho} F^{\sigma \rho}\right)^{q}\Big],
\end{equation*}
\end{subequations}
finding an asymptotically AdS black hole solution for $\xi$ given in (\ref{eq:rangexi}), where the line element takes the form
\begin{eqnarray}
ds^2&=&-N^2(r)F(r) dt^2+\frac{dr^2}{F(r)}+r^2 \sum_{i=1}^{D-2} dx_{i}^{2},\label{metric-sol}\\
N(r)&=&1,\qquad F(r)=r^2\big[1-( \eta \Phi(r)^{2})^{\frac{1}{n-1}}\big],\label{metric-function-sol}
\end{eqnarray}
with the scalar field $\Phi(r)$ given by (\ref{eq:phixi}), and the Maxwell strength reads
\begin{eqnarray}\label{eq:Frt}
F_{rt}&=&(A_{t})'=\frac{Q}{\left(r^{D-2} \epsilon(\Phi)\right)^{\frac{1}{2q-1}}},
\end{eqnarray}
and the integration constants $Q$ and $a$ are tied by the relation
\begin{eqnarray}
\label{relcst}
Q = \left[\dfrac{4(\eta - \xi)\xi \eta^{\frac{1}{n-1}}}
{q (-2)^{q} (n-1)^2(1-4\xi)^2 a^{\frac{2(D-2)q}{2q-1}}}\right]^{\frac{1}{2q}}.
\end{eqnarray}
\begin{table}
\caption{\label{tabla1} Admissible values for the parameters $\eta,q$ ensuring a real solution.}
\begin{tabular}{|c|c|c|c|}
\hline
Sign of $q$ & num($q$) & denom($q$) & Range for $\eta$\\
\hline \hline Case 1: $q>0$ & even & odd & $\eta \geq \xi$\\
[1ex] \hline \hline Case 2: $q>0$ &odd & odd & $0\leq \eta \leq \xi$\\ 
[1ex] \hline \hline Case 3: $q<0$ &even & odd & $0\leq \eta \leq \xi$\\
[1ex] \hline \hline Case 4: $q<0$ &odd & odd & $\eta \geq \xi$\\
[1ex] \hline
\end{tabular}
\end{table}
A few comments can be made at this point. First, we note that the Ricci scalar for (\ref{metric-sol})-(\ref{metric-function-sol}), with the scalar field given previously in (\ref{eq:phixi}), reads
\begin{eqnarray*}
R&=&\left[D(D-1)+\frac{2 (\Phi''r+2D\Phi')r}{\Phi(n-1)}-\frac{2(n-3)(\Phi')^2r^2}{(n-1)^2 \Phi^2}\right]\\
&\times&{(\eta \Phi^2)^{\frac{1}{n-1}}}-{D(D-1)},
\end{eqnarray*}
showing the existence of a curvature singularity located at
$$r_{s}=\frac{b}{a}.$$
To complement the above, the coordinate singularity $F(r)= 0$ {leads to a sign restriction on $\eta$. To avoid naked singularities, hereafter we will assume $\eta > 0$}, and as a consequence, the curvature singularity $r_s$ is always hidden by an event horizon located at 
\begin{equation}\label{eq:rh}
r_h=\frac{b}{a}+\frac{\eta^{\frac{1-4\xi}{4\xi}}}{a}.
\end{equation} 
On second place, it is observed that (\ref{relcst}) imposes extra restrictions in our parameters $\eta,q$. Indeed, since we are interested in real solutions, the constraint between $Q$ and $a$ forces $\dfrac{\eta - \xi}{q(-2)^q}$ to be non negative. The allowed configurations are shown in the Table \ref{tabla1}.

 {Here is important to note that the inclusion of the Maxwell field (\ref{eq:maxwell}), given by  (\ref{eq:Frt})-(\ref{relcst}), allow us to obtain a new hairy charged black hole solution where the location of the event horizon $r_h$ (\ref{eq:rh}) is written in function of the parameter $\eta$. Indeed, one can interpret the role of this parameter by saying that it generates a shift in the location of $r_h$. In addition, it is worth pointing out the importance of the Maxwell strength (\ref{eq:Frt})-(\ref{relcst}) in order to obtain non-null thermodynamical quantities. In fact, given that we are not adding a gravity theory contribution, {\em{a priori}} the entropy ${\mathcal{S}}$ keep being proportional to (\ref{Sprop}), where now with the new location of the event horizon (\ref{eq:rh}) 
$${\mathcal{S}} \propto (\eta-\xi), $$
while that the electric charge becomes
$${\mathcal{Q}}_{e} \propto (\eta-\xi)^{\frac{2q-1}{2q}},$$
where the limit $\eta \to \xi$ makes the Maxwell source term to vanish, returning exactly to the hairy solution found previously in \cite{Correa:2013bza}.
Finally,} it is observed that the coupling function $\epsilon(\Phi)$ (\ref{eq:epsilon}) is positive outside the singularity provided that $4(1+(n-1)D)\xi - (n-1)(D-1) \geq 0$, imposing a bound on the nonminimal coupling $\xi$. All of these facts will be fully explored and discused in the following section.

We finish this section briefly reporting the solution for the case $\xi=1/4$. Here the potential $U_{\eta,q}(\Phi)$ as well as the coupling function $\epsilon(\Phi)$ read
\begin{eqnarray*}
U_{\eta,q}(\Phi)&=&\frac{\Phi^2}{8} \Big[4\, \ln  \left( {\frac {\Phi}{b}}
 \right)^{2}+ 4\left( D-1 \right) \ln  \left( {\frac {\Phi
}{b}} \right)\nonumber\\
 &+& \left( D-1 \right)  \left( D-2 \right)  \Big]-\frac{ \eta^{\frac{1}{n-1}}\Phi^{\frac{2n}{n-1}}}{8 q n (n-1)^2}\nonumber\\
 &\times& \Big[{n}^{2} \left( 8\eta -2-4q(4 \eta-n) \right) \ln  \left( {\frac
{\Phi}{b}} \right)^{2}\\
&+&n \left( n-1 \right)  \big(D(4\eta-1)-4q(4\eta-n)(D-1)\big) \\
&\times& \ln
 \left( {\frac {\Phi}{b}} \right) + q\left( n-1 \right) ^{2} \left( D-1
 \right)  \left( D-2 \right) \\
 &\times& \left( n-4\,\eta \right)\Big],
\\ \\
\epsilon(\Phi)^{\frac{1}{2q-1}}&=&\frac{\Phi^{-\frac{2n}{n-1}}}{D(n-1)+2n \ln\left(\frac{\Phi}{b}\right)}\,\ln\left(\frac{\Phi}{b}\right)^{\frac{1+2(1-D)q}{2q-1}},
\end{eqnarray*}
while that the metric is given by (\ref{metric-sol})-(\ref{metric-function-sol}) where now the scalar field $\Phi$ acquires the structure found previously in (\ref{eq:phixi14}), and the Maxwell tensor is represented by (\ref{eq:Frt}) but now the integration constants $Q$ and $a$ are related as
$$Q=\left[\frac{(4\eta-1)\eta^{\frac{1}{n-1}}}
{2 q(-2)^q a^{\frac{2q(D-2)}{2q-1}} (n-1)^2}\right]^{\frac{1}{2q}}.$$
After this analysis for this electrically charged hairy black hole solution, now we will study the thermodynamics for the case $0<\xi<\frac{1}{4}$ through the Euclidean action.

\section{Thermodynamics of the charged hairy black hole solution}\label{Section-term}

We now turn into the study of the themodynamic quantities of the solution (\ref{metric-sol})-(\ref{metric-function-sol}) by means of the Euclidean approach, where the partition function is identified with the Euclidean path integral in the saddle point approximation around the Euclidean continuation of the classical solution \cite{Gibbons:1976ue, Regge:1974zd}, where the time coordinate $\tau=it$ is imaginary and periodic with period $\beta = T^{-1}$. The Euclidean action is related to the Gibbs free energy $\mathcal{G}$ by
\begin{equation}\label{eq:gibbs}
I_{E} = \beta \mathcal{G} = \beta (\mathcal{M} - T\mathcal{S} - \Phi_e \mathcal{Q}_e),
\end{equation}
where $\mathcal{M}$ is the mass, $\mathcal{S}$ is the entropy and $\Phi_e,\mathcal{Q}_e$ stands for the electric potential and electric charge, respectively. For this analysis, it is enough to consider the following class of static metrics $$ds^2 = N(r)^2F(r)d\tau^2 + \dfrac{dr^2}{F(r)} + r^2d\Sigma_{D-2}^2.$$ Thus,  the reduced Euclidean action reads
\begin{eqnarray}
I_{E} &=& \beta  \Sigma_{D-2}  \int_{r_h}^{\infty} dr [ N\mathcal{H} + A_{\tau}p']  + B,\label{eq:red-act-1}\\
\mathcal{H}&=&-\frac{1}{2n}(D-2)\frac{d}{dr}
\left[r^{D-1}\left(1-\frac{F}{r^2}\right)^n\right]\nonumber\\
&+&
r^{D-2}\Big\{\left(\frac{1-4\xi}{2}\right)F(\Phi^{\prime})^2-
\left(F^{\prime}+\frac{2(D-2)}{r}F\right)\nonumber\\
&\times&\xi\Phi\Phi^{\prime}+\Phi^2\left(-\frac{\xi}{2r^2}(D-2)(D-3)F\right)\nonumber\\
&-&2\xi\Phi\Phi^{\prime\prime}F\nonumber
-\frac{(D-2)\xi}{2r}F^{\prime}\Phi^2+U_{\eta,q}(\Phi) \Big\}\nonumber\\
&+&\frac{(2q-1)}{2q} \left(\frac{p^{2q}}{ q (-2)^{q-1} r^{D-2} \epsilon(\Phi)}\right)^{\frac{1}{2q-1}}, \label{eq:red-act-2}
\end{eqnarray}
where $\tau \in [0,\beta]$, $r\geq r_{h}$, $\Sigma_{D-2}$ represents the volume of the compactified $(D-2)$- dimensional euclidean manifold, and $B$ is a boundary term, which is properly fixed by requiring that the reduced action has an extremum. In this reduced action,
$$p = {q (-2)^{q-1} r^{D-2}\epsilon(\Phi) }\left(\frac{A_{\tau}'}{N}\right)^{2q-1},$$ is the conjugate momentum of $A_{\tau}$, and by Gauss Law, $p = constant \equiv {Q}_e$. We work in the grand canonical ensemble, so that we will consider variations of the action where the temperature $T$ and the potential $\displaystyle \Phi_e \equiv \lim_{r \rightarrow +\infty}A_{\tau}(r) - A_{\tau}(r_h)$ are fixed. In order to avoid divergent terms, we must restrict the admissible values of $\xi$ to the range:
\begin{eqnarray}
\label{xi:bound2} \xi^{\star} \equiv \dfrac{(n-1)(D-1)}{4(1+(n-1)D)}\leq \xi<\frac{1}{4},
\end{eqnarray}
leading to the following electric potential:
\begin{eqnarray}\label{eq:pot}
\Phi_e&=& \frac{r_h (n-1)^{\frac{q-1}{q}}(1-4\xi)^{\frac{q-1}{q}}}{b+\eta^{\frac{1-4\xi}{4\xi}}} \left(\dfrac{4(\eta-\xi)\xi\eta^{\frac{1}{n-1}}}{q(-2)^q} \right)^{\frac{1}{2q}} \nonumber\\
&\times& \Big[
\eta^{-\frac{4\xi+n-1}{4 (n-1)\xi}}\left(\eta^{\frac{1-4\xi}{4\xi}}+b\right)^D-\delta_{\xi}^{\xi^{\star}}\Big],\end{eqnarray}
where
$$\delta_{\xi}^{\xi^{\star}} =   \begin{cases}
1&,\quad \xi = \xi^{\star}, \\
\ \\
0&,\quad \xi \neq \xi^{\star}, \\
\end{cases}$$ 
and now $r_h$ is the location of the event horizon given in (\ref{eq:rh}). For the black hole solution (\ref{metric-function-sol}), the Hawking temperature $T$ reads
\begin{eqnarray}
T = \dfrac{(1+b\eta^{\frac{4\xi-1}{4\xi}}) r_h \xi}{(n-1)(1-4\xi)\pi} = \beta^{-1}.
\end{eqnarray}

With all above, the variation of the reduced action leads to
\begin{eqnarray*}
\delta B &=& \beta \Sigma_{D-2}\left[ \left\lbrace - \dfrac{1}{2}N(D-2)r^{D-3}\left(1 - \dfrac{F}{r^2}\right)^{n-1} \right. \right. \\
&+&\left. Nr^{D-2}\xi\Phi\Phi' + \dfrac{1}{2}(D-2)\xi\Phi^2r^{D-3} \right\rbrace \delta F   \\
&+&\left. 2Nr^{D-2}\xi\Phi F \delta \Phi'-2Nr^{D-2}\Big(\dfrac{F\Phi'}{2}- \xi\Phi'F \right.\nonumber\\
&+& \dfrac{\xi\Phi F'}{2} \Big)\delta\Phi -\left.A_{\tau}\delta p \right],
\end{eqnarray*}
where this variation has to be computed at $r=+\infty$ and at the event horizon $r=r_{h}$. At the infinity, we obtain
\begin{align*}
&\delta B(\infty) = \beta\Sigma_{D-2} \\
&\displaystyle \times \lim_{r\to \infty} \left(-\dfrac{2(D-2)\xi \eta^{\frac{1}{n-1}}(\eta - \xi) r^{D}\Phi(r)^{\frac{4\xi+n-1}{2\xi(n-1)}}}{(1-4\xi)(n-1)} \right)\delta a,
\end{align*}
where it can be observed that this limit is divergent unless $\eta = \xi$ or $\xi^{\star}\leq \xi<1/4$, as in (\ref{xi:bound2}). Additionally, this lower bound in the latter condition determines the unique value for $\xi$ that ensures a finite and nonzero contribution to the boundary term. In that specific case, we obtain $$\delta B(\infty) = -\beta\Sigma_{D-2}\cdot \left(\dfrac{2(D-2)\xi^{\star} \eta^{\frac{1}{n-1}}(\eta - \xi^{\star})}{(1-4\xi^{\star})(n-1)a^D}\right) ~ \delta a.$$
In order to compute the variation at the horizon, we use the relations
\begin{eqnarray*}
\delta F\Big{|}_{r_h} &=& -F'(r_{h})~\delta r_h,\quad \delta p = \delta Q_e,\\
\delta \Phi \Big{|}_{r_h} &=& \delta(\Phi(r_{h})) - \Phi'(r_h)\delta r_{h},
\end{eqnarray*}
and a simple calculation leads to
$$\delta B(r_h)= 2\pi \Sigma_{D-2} \left(1 - \dfrac{\xi}{\eta} \right)~ \delta r_{h}^{D-2} + \beta \Sigma_{D-2}\Phi_e\delta Q_e.$$
Thus, the boundary term is
\begin{eqnarray}\label{Bound}
B &= \beta\Sigma_{D-2} \dfrac{2\xi\eta^{\frac{1}{n-1}}(\eta - \xi)(D-2)}{(D-1)(n-1)(1-4\xi)a^{D-1}}~\delta_{\xi}^{\xi^{\star}} \nonumber \\
&- 2\pi\Sigma_{D-2}\left(1-\dfrac{\xi}{\eta}\right)r_{h}^{D-2} - \beta\Sigma_{D-2}\Phi_e \mathcal{Q}_e,
\end{eqnarray}
and the identification of the thermodynamics parameters is obtained by comparing (\ref{Bound}) with (\ref{eq:gibbs}), where the thermodynamic quantities read:
\begin{align}
\mathcal{M} &= \Sigma_{D-2}\left(\dfrac{2\xi\eta^{\frac{1}{n-1}}(\eta - \xi)(D-2)}{(D-1)(n-1)(1-4\xi)a^{D-1}}\right)~\delta_{\xi}^{\xi^{\star}}  \nonumber\\
&=\Sigma_{D-2}\left(\dfrac{2\xi\eta^{\frac{1}{n-1}}(\eta - \xi)(D-2)}{(D-1)(n-1)(1-4\xi)}\right) \nonumber\\
&\times\left(\frac{r_h}{b+\eta^{\frac{1-4\xi}{4\xi}}}\right)^{D-1}~\delta_{\xi}^{\xi^{\star}},\label{eq:mass}\\
\mathcal{S} &= 2\pi\Sigma_{D-2}\left(1-\dfrac{\xi}{\eta}\right)r_{h}^{D-2}, \label{eq:entrop}\\
\mathcal{Q}_e &= {\Sigma_{D-2} Q_e} \nonumber\\
&=\dfrac{\Sigma_{D-2}~q(-2)^{q-1}}{a^{D-2}}\left[\dfrac{4(\eta - \xi)\xi \eta^{\frac{1}{n-1}}}
{q (-2)^{q} (n-1)^2(1-4\xi)^2 }\right]^{\frac{2q-1}{2q}} \nonumber\\
&={\Sigma_{D-2}~q(-2)^{q-1}} \left(\frac{r_h}{b+\eta^{\frac{1-4\xi}{4\xi}}}\right)^{D-2} \nonumber\\
&\times \left[\dfrac{4(\eta - \xi)\xi \eta^{\frac{1}{n-1}}}
{q (-2)^{q} (n-1)^2(1-4\xi)^2 }\right]^{\frac{2q-1}{2q}}. \label{eq:charge}
\end{align}
It is a simple exercise to check that the First Law of Black Holes Thermodynamics $$\delta \mathcal{M} = T\delta \mathcal{S} + \Phi_e \delta \mathcal{Q}_e,$$
holds.

Following the procedure performed in \cite{Banados}, we note that the reduced action (\ref{eq:red-act-1})-(\ref{eq:red-act-2}) has the scaling symmetries
\begin{eqnarray*}
\bar{r}&=&\sigma r,\quad \bar{N}(\bar{r})=\sigma^{1-D} N(r),\quad \bar{F}(\bar{r})=\sigma^{2} F(r),\\
\bar{\Phi}(\bar{r})&=&\Phi(r),\quad \bar{p}(\bar{r})=\sigma^{D-2} p(r),\quad \bar{A}_{\tau}(\bar{r})=\sigma^{2-D} A_{\tau}(r),
\end{eqnarray*}
where $\sigma$ is a positive constant, allowing a Noether current
\begin{eqnarray*}
C(r) &=& \left[ \left\lbrace - \dfrac{1}{2}N(D-2)r^{D-3}\left(1 - \dfrac{F}{r^2}\right)^{n-1} \right. \right. \\
&+&\left. Nr^{D-2}\xi\Phi\Phi' + \dfrac{1}{2}(D-2)\xi\Phi^2r^{D-3} \right\rbrace (-rF'+2F)   \\
&+&\left. 2Nr^{D-2}\xi\Phi F (-r\Phi''-\Phi')-2Nr^{D-2}\Big(\dfrac{F\Phi'}{2} \right.\nonumber\\
&-& \xi\Phi'F+ \dfrac{\xi\Phi F'}{2} \Big)(-r\Phi') -\left.A_{\tau} \big(-rp'+(D-2)p\big) \right],
\end{eqnarray*}
which is conserved ($C'(r)=0$). Evaluating at infinity and at the horizon $r_h$:
\begin{eqnarray*}
C(\infty)&=&\frac{\mathcal{M} (D-1)}{\Sigma_{D-2}},\\
 C(r_h)&=&\frac{T \mathcal{S} (D-2)}{\Sigma_{D-2}}+\frac{\Phi_e \mathcal{Q}_{e} (D-2)}{\Sigma_{D-2}},
\end{eqnarray*}
and given that $C(r)$ is a constant, we have $C(\infty)=C(r_h)$, permitting to obtain  a $D$-dimensional Smarr relation \cite{Smarr:1972kt}
\begin{eqnarray}\label{eq:smarr}
\mathcal{M}= \left(\dfrac{D-2}{D-1}\right) ( T\mathcal{S} + \Phi_e\mathcal{Q}_e).
\end{eqnarray}

Many commentaries can be made from these thermodynamical quantities. {Restoring $\kappa = 8\pi G$, it can be observed that the entropy satisfies $\mathcal{S} = \dfrac{\mathcal{A}}{4\tilde{G}}$, where $$\tilde{G} = \dfrac{G}{1-\frac{\xi}{\eta}}$$ is an ``effective'' Newton constant. If one imposes the positivity of this effective Newton constant, then $\eta > \xi$, concluding that these configurations only are feasible for Cases 1 or 4 from Table \ref{tabla1}.} 

{Together with the above, for the uncharged case (when $\eta \rightarrow \xi$) all the thermodynamical quantities vanish except for the Hawking Temperature, where 
$$T = \dfrac{(1+b\xi^{\frac{4\xi-1}{4\xi}}) r_h \xi}{(n-1)(1-4\xi)\pi} \neq 0, $$
recovering the analisys performed in \cite{Correa:2013bza}.
}  

\begin{figure}[!ht]
\begin{center}
\includegraphics[scale=0.15]{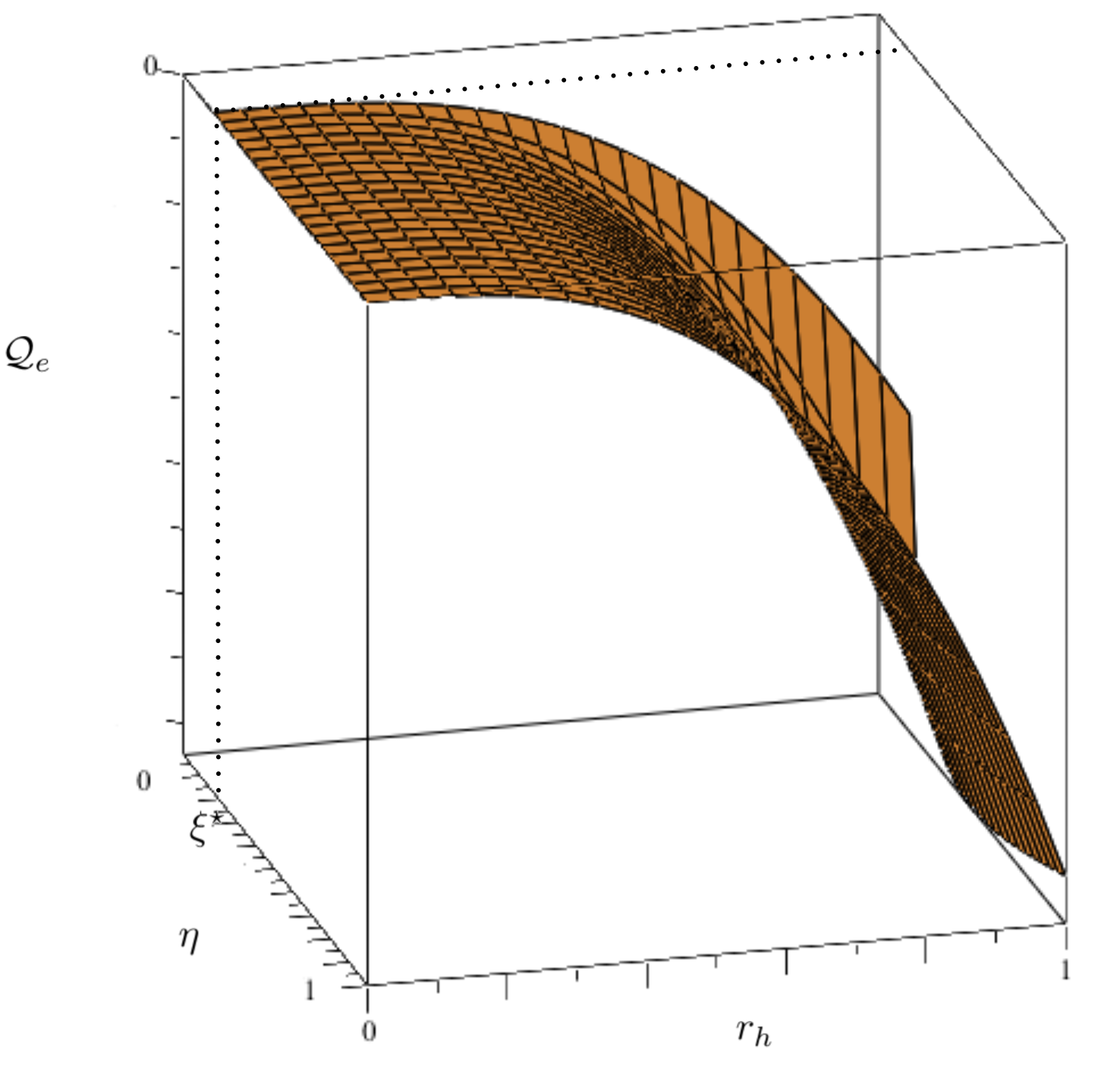}
\includegraphics[scale=0.27]{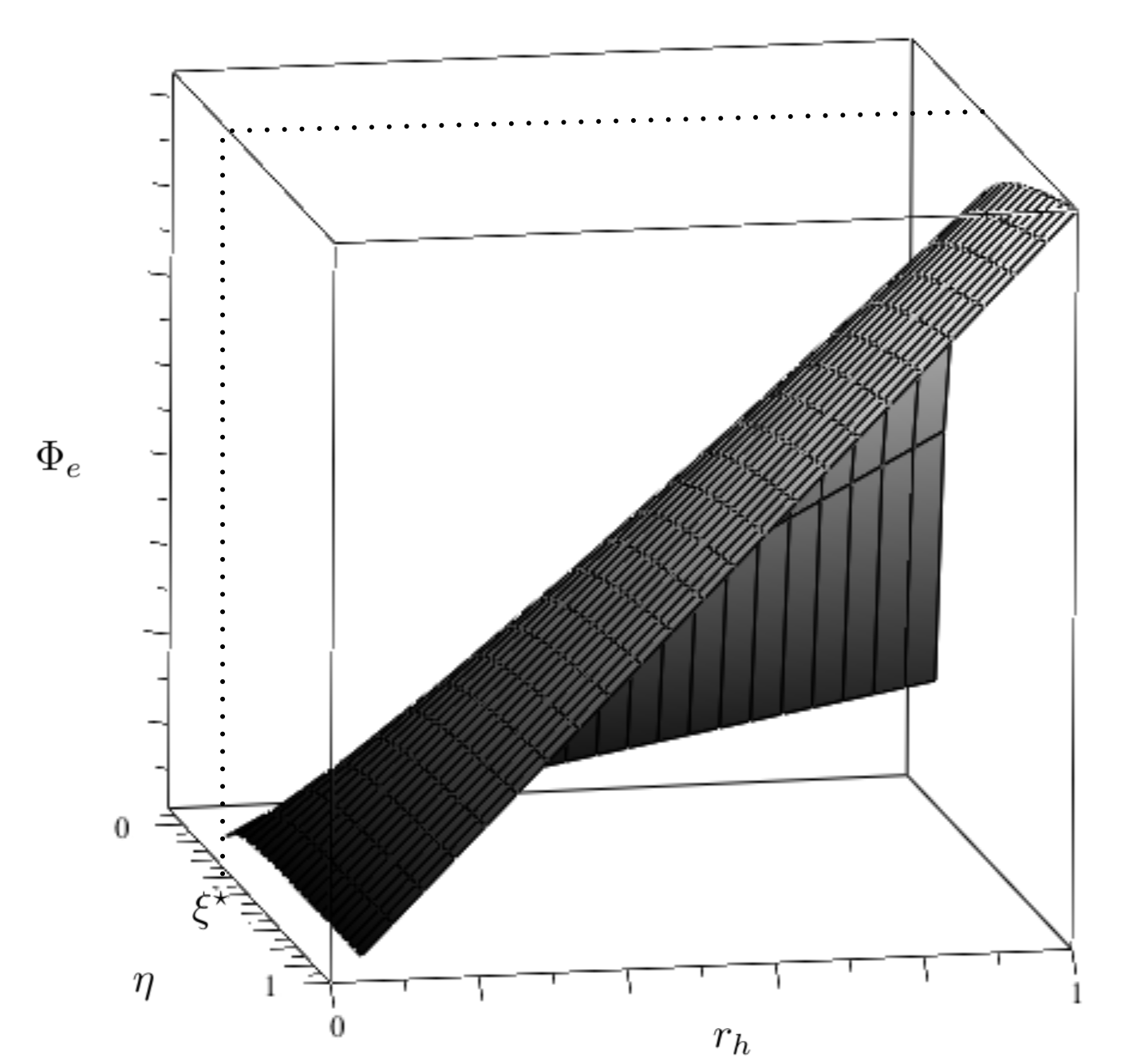}
\caption{Up panel: Electric charge $\mathcal{Q}_{e}$ versus the parameter $\eta$ and the location of the event horizon $r_h$, which is a negative quantity. Down panel: Electric potential $\Phi_{e}$,  versus the parameter $\eta$ and the location of the event horizon $r_h$, which is a positive quantity. For both cases, we suppose $0<r_h<1$ and $0<\eta<1$. }
\label{p1}
\end{center}
\end{figure}

Curiously enough, from equations (\ref{eq:charge}) and (\ref{eq:pot}) we have that the electric charge and electric potential have opposite signs, as shown in Figure \ref{p1}. The above is due to we are working on real solutions, so $(\eta-\xi)/\big[q (-2)^q\big]$ is a non-negative quantity, as we had already noted. This leads to $\mathcal{Q}_{e}<0$. In what respects to $\Phi_{e}$, it can be noticed that the expression
\begin{equation}\label{eq:cond}
\eta^{-\frac{4\xi+n-1}{4 (n-1)\xi}}\left(\eta^{\frac{1-4\xi}{4\xi}}+b\right)^D-\delta_{\xi}^{\xi^{\star}}
\end{equation} 
is always positive for $\xi^{\star}\leq\xi<1/4$, leading to $\Phi_{e}>0$. The above is reinforced by the fact that through the Smarr formula (\ref{eq:smarr}) 
$$0 < \mathcal{M}<\left(\frac{D-2}{D-1}\right)T \mathcal{S},$$ implying that $\Phi _e\mathcal{Q}_e < 0$.

It is worth pointing out that with these thermodynamical quantities we can study the local stability considering small perturbation around the equilibrium. As a first step, we rewrite the temperature $T$ and the electric potential $\Phi_{e}$, in the function of the extensive quantities $\mathcal{S}$ and $\mathcal{Q}_{e}$,  given by
\begin{eqnarray*}
T&=&\displaystyle{\frac{\left( {\eta}^{-{\frac {4\,\xi-1}{4\xi}}}+b \right) \xi\,{\eta}^{
{\frac {4\,\xi-1}{4 \xi}}} \left( {\frac {\mathcal{S}\eta}{2 \pi \,\Sigma_
{D-2} \left( \eta-\xi \right) }} \right) ^{ \frac{1}{ D-2}}
}{(n-1)(1-4\xi)\pi}}, \\
\Phi_{e}&=&  \frac{\left( 1-4\,\xi \right)  \left( n-1 \right)} {\left({
\frac {\Sigma_{D-2} \left( -2 \right) ^{q-1}q}{\mathcal{Q}_e}} \right) ^{\frac{1}{
  D-2 } }} \\
&\times&\left[{\frac {4 \xi\, \left( \eta-\xi \right) {\eta}^{ \frac{1}{ n-1}}}{ \left( -2 \right) ^{q}q \left( 1-4\,\xi \right) ^{2} \left( n
-1 \right) ^{2}}}
\right]^{\frac{D-2q-1}{2(D-2)q}}\\
&\times&\Big[
\eta^{-\frac{4\xi+n-1}{4 (n-1)\xi}}\left(\eta^{\frac{1-4\xi}{4\xi}}+b\right)^D-\delta_{\xi}^{\xi^{\star}}\Big],
\end{eqnarray*}
and following the steps performed in \cite{Mansoori:2013pna,Mansoori:2014oia}, the specific heat $C_{\Phi_{e}}$ at constant electrical potential as well as the electric permittivity $\epsilon_{T}$ at constant temperature  read 
\begin{eqnarray*}\label{eq:cq}
C_{\Phi_{e}}&\equiv& T\left(\frac{\partial \mathcal{S}}{\partial T}\right)_{\Phi_{e}}=\frac{T \{\mathcal{S},\Phi_{e}\}_{\mathcal{S},\mathcal{Q}_{e}}}{\{T,\Phi_{e}\}_{\mathcal{S},\mathcal{Q}_{e}}}\\
&=&(D-2)\mathcal{S}\cdot \delta_{\xi}^{\xi^{\star}},\\ \\
\epsilon_{T}&\equiv&\left(\frac{\partial \mathcal{Q}_{e}}{\partial \Phi_{e}}\right)_{T}=\frac{\{\mathcal{Q}_{e},T\}_{\mathcal{S},\mathcal{Q}_{e}}}
{\{\Phi_{e},T\}_{\mathcal{S},\mathcal{Q}_{e}}}\\
&=&\frac{(D-2) \Sigma_{D-2} \mathcal{Q}_{e}}{(n-1) (1-4 \xi)} \left(\frac{\Sigma_{D-2} (-2)^{q-1} q}{\mathcal{Q}_{e}}\right)^{\frac{1}{D-2}}
\\
&\times&\frac{\left[\frac{4 \xi (\eta-\xi)\eta^{\frac{1}{n-1}}}{q (-2)^{q} (1-4\xi)^2 (n-1)^2}\right]^{\frac{2q+1-D}{2(D-2)q}}}{\eta^{\frac{1-4\xi-n}{4\xi (n-1)}}
\left(\eta^{\frac{1-4\xi}{4\xi}}+b\right)^{D}-{\delta_{\xi}^{\xi^{\star}}}},
\end{eqnarray*} 
where we are considering the partial derivatives in the function of Poisson brackets, where if $f$ and $g$ are explicit functions of $a$ and $b$: 
$$\{f,g\}_{a,b}:=\left(\frac{\partial f }{\partial a}\right)_{b}\left(\frac{\partial g} {\partial b}\right)_{a}
-\left(\frac{\partial f} {\partial b}\right)_{a}\left(\frac{\partial g} {\partial a}\right)_{b}.$$
The case $\xi=\xi^{\star}$ deserves to be discussed. The specific heat $C_{\Phi_{e}}$ is now positive for $\eta>\xi^{\star}$, represented again by Cases 1 and 4  from  Table \ref{tabla1} and interpreted as local stability under thermal fluctuations. Nevertheless, with respect to the electrical permittivity $\epsilon_{T}$, this is negative, being an unstable configuration under electrical perturbations. This is not surprising, since this hairy electrical configuration has opposite signs between the charge $\mathcal{Q}_{e}$ and the electrical potential $\Phi_e$, being explained by the fact that  $\mathcal{Q}_{e}$ increases when $\Phi_e$ decreases and the system leave the equilibrium state \cite{Gonzalez:2009nn,Chamblin:1999hg}. 

\begin{figure}[!ht]
\begin{center}
\includegraphics[scale=0.1]{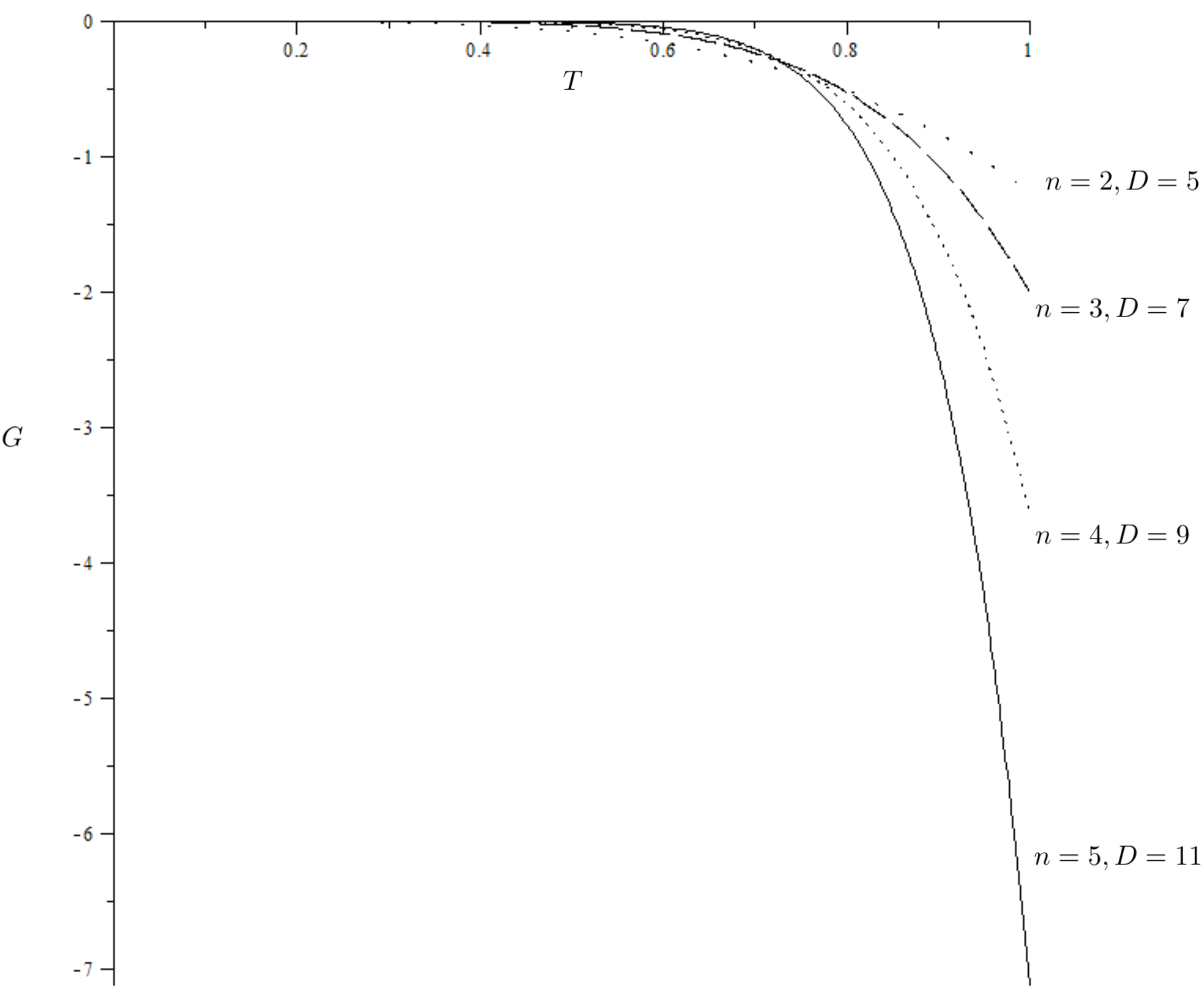}
\caption{Gibbs free energy $\mathcal{G}$ versus temperature $T$ for some particular cases. Here we note that when the dimension $D$ of the solution and the integer $n$ increase, $\mathcal{G}$ becomes more negative.}
\label{gibbs}
\end{center}
\end{figure}

Finally, the Global stability can be studied in the grand canonical ensemble with the Gibbs free energy $\mathcal{G}=I_{E}/\beta$ as a state function, wherein our case and by using the Smarr relation (\ref{eq:smarr}) is given by
\begin{eqnarray}
\mathcal{G}&=&\mathcal{M}-T \mathcal{S}-\Phi_e \mathcal{Q}_{e}=-\frac{\mathcal{M}}{D-2},\nonumber\\
\label{eq:Gibbs} 
\end{eqnarray}
which is negative. For these black hole configurations, the concavity condition simply reads
\begin{eqnarray*}
\frac{\partial^{2}\mathcal{G}}{\partial T^2}&=&- \dfrac{C_{\Phi_e}}{T},
\end{eqnarray*}
ensuring global thermodynamic stability (${\partial^{2}\mathcal{G}}/{\partial T^2}<0$). Some particular cases are presented in Figure \ref{gibbs}.

\section{Conclusions and discussions}\label{Section-conclusions}

In the present paper, we extend the model presented in \cite{Correa:2013bza}, characterized by a special truncation of the Lanczos-Lovelock gravity theories dressed by a scalar field non-minimal coupling together with a suitable choice of the potential. To perform this, we add a special matter source characterized by a non-linear Maxwell field coupling with a function depending on the scalar field. From this model, there is the presence that an integration constant that for our situation no longer a hair, making possible a nontrivial thermodynamics analysis for the black hole solutions in a certain range for the non-minimal coupling $\xi$, exhibiting a limit in the proposed theory that recovers the original result from \cite{Correa:2013bza}.
 
Computing its thermodynamics parameters, by using the Euclidean action, we obtain interesting thermodynamical quantities satisfying the higher dimensional First Law of Thermodynamics. From this regularized action and thanks to its scaling symmetries, we obtain a Noether current allowing us a derivation of a Smarr formula, following the procedure elaborated in \cite{Banados}.  It is worth pointing out that the mass is the only null quantity except for a special election for the non-minimal coupling parameter $\xi$. The above allows us to analyze local and global thermodynamical stability for some election of the coupling parameter, following the criteria from the non-negativity of the specific heat $C_{\Phi_{e}}$ as well as the electrical permittivity $\epsilon_{T}$, showing that this charged hairy higher dimensional configuration is locally stable under thermal fluctuations but is unstable under electrical fluctuations, due in part to the opposite signs present between the electrical charge and the electrical potential. Supplementary to the above, the global stability by using the concavity criteria for the Gibbs free energy, is assured only if $\xi=\xi^{\star}$.

Just for completeness, the case $b=0$ is briefly discussed. When $\xi \neq \xi^{\star}$, and the coupling function $\epsilon (\Phi)$ (\ref{eq:epsilon}) can be set to a constant, provided the following relation between $q$ and $\xi$: $$\dfrac{4\xi n}{(n-1)(1-4\xi)} = \dfrac{2(D-2)q}{2q-1}.$$ On contrast, the case $b=0$ and $\xi = \xi^{\star}$ recovers the stealth configuration already found in \cite{Gaete:2013oda,Gaete:2013ixa}, which is a particular solution of the Einstein equations with the property that the gravity and matter source section vanish identically, this is
$${{\cal G}}^{(n)}_{\mu\nu}=0=T_{\mu \nu}^{\Phi}.$$

Some natural extensions of this work would be for example to consider the inclusion of matter sources such as linear Maxwell fields \cite{BravoGaete:2019rci} to obtain electric and/or magnetic configurations, or the addition of other non-linear electrodynamics theories (see for example \cite{Plebanski:1968,Alvarez:2014pra,Stetsko:2020nxb,Stetsko:2020tjg}).

\begin{acknowledgments}
 We would
like to thank especially to Mokhtar Hassaine for stimulating discussions and nice comments
to improve the draft. MB wishes to dedicate this work in memory of his mentor and friend
Gonzalo Hidalgo Am\'estica. SG is part of the research group GEMA Res. 180/2019 VRIP-UA.
\end{acknowledgments}

\section{Appendix}\label{Section-Appendix}

In this Section, we report the equations of motions (\ref{eq:gmunu}) varying the action (\ref{eq:actioncorrea}) respect to the metric $g_{\mu \nu}$.  Each one is given by

\begin{widetext}
\begin{eqnarray*}
{\cal{G}}^{(n)t}_{\,\,t}&=&{\cal{G}}^{(n)r}_{\,\,r}=-\frac{(D-2)}{2n}\,\left(nr f'+(D-1)f\right) f^{n-1},\\ \\
{\cal{G}}^{(n)x_{i}}_{\,\,x_{i}}&=&-\frac{f^{n-2}}{2n}
\big(n r^2 f''f+n(n-1)r^2 (f')^2+2n f r (D-1)f'\nonumber\\
&+&(D-2)(D-1)f^2\big),\\ \\
{T}^{\Phi t}_{\,\,t}&=& 2\xi r^2 \Phi (1-f) \Phi''+2(1-f)r^2\left(\xi-\frac{1}{4}\right) (\Phi')^2\nonumber\\
&-&2 \left[\frac{1}{2}r
f'-(D-1)(1-f)\right] r\xi \Phi
\Phi'-\frac{1}{2} r \xi \Phi^2 (D-2)
f'\nonumber\\
&+&\frac{1}{2} (D-1) (D-2) \xi(1-f)\Phi^2-U(\Phi), \\ \\
{T}^{\Phi r}_{\,\,r}&=& \frac{1}{2}r^2(1-f)
(\Phi')^2-2\left[\frac{1}{2} r
f'-(D-1)(1-f)\right] r \xi \Phi \Phi'
\\
&-&\frac{1}{2} r \xi \Phi^2 (D-2)
f'+\frac{1}{2}(D-1)(D-2) \xi(1-f)\Phi^2\\
&-&U(\Phi), \\ \\
{T}^{\Phi x_i}_{\,\,x_i}&=&2\xi r^2 \Phi(1-f)
\Phi''-\frac{1}{2}r^2 f'' \xi \Phi^2
+2(1-f)r^2 \left(\xi-\frac{1}{4}\right)(\Phi')^2\\
&-&2 \left[r f'-(D-1)(1-f)\right] r \Phi \xi
\Phi'-r \xi \Phi^2 (D-1) f' \\
&+&\frac{1}{2}(D-1)(D-2) \xi(1-f)\Phi^2\\
&-&U(\Phi),
\end{eqnarray*}
while that the potential $U(\Phi)$ in \cite{Correa:2013bza} reads:
\begin{eqnarray*}
U(\Phi)&=& \frac{1}{(1-4\xi)^2}\sum_{i=1}^{6} \alpha_{i} \Phi^{\gamma_{i}},
\end{eqnarray*}
with
\begin{eqnarray*}
\alpha_{1}&=&\frac{[4(D-1)\xi-D+2]
[4\xi D-D+1]\xi}{2},\qquad \gamma_1=2,\\ \\
\alpha_{2}&=&{4[4(D-1)\xi-D+2]b\xi^2},\qquad \gamma_2=\frac{1}{2 \xi},\\ \\
\alpha_3&=&{2\xi^2 b^2},\qquad \gamma_3=\frac{1-2\xi}{\xi},\\ \\
\alpha_4&=& -\dfrac{\xi ^{\frac{n}{n-1}}}{2n(n-1)}\, [4 n D \xi-(n+4\xi-1)(D-1)][4\xi((n-1)D-n+2)-(n-1)(D-2)] ,\quad \gamma_4=\frac{2n}{n-1},\\ \\
\alpha_5&=& \dfrac{16b\xi(n(D-1)-(D-2))-4D(n-1)+8(n-1)}{n-1}\xi^{\frac{2n-1}{n-1}},\qquad \gamma_5=\frac{4 \xi+n-1}{2 \xi (n-1)}, \\ \\
\alpha_6&=& -\frac{2\xi^{\frac{2n-1}{n-1}}b^2(n+4\xi-1)}{(n-1)},\qquad \gamma_6=\frac{(4-2n)\xi+n-1}{\xi(n-1)}.\\
\end{eqnarray*}
\end{widetext}


\end{document}